\documentclass[aps,prb,twocolumn,superscriptaddress,
               amsmath]{revtex4}

\usepackage{graphicx}
\usepackage{dcolumn}
\usepackage{bm}
\usepackage{natbib}

\newcommand{\eqn}[1]{\vspace{-0.cm}\begin{equation}
#1
\end{equation}}

\renewcommand{\=}{\,=\,}
\newcommand{\+}{\,+\,}
\renewcommand{\-}{\,-\,}

\newcommand{\trip}[3]{\ensuremath{\left[T_{#1},\left[T_{#2},T_{#3}\right]\right]}}

\newcommand{\f}[2]{{\ensuremath{\mathchoice%
        {\dfrac{#1}{#2}}
        {\dfrac{#1}{#2}}
        {\frac{#1}{#2}}
        {\frac{#1}{#2}}
        }}}

\newcommand{\ket}[1]{\ensuremath{\left| #1 \right\rangle}}
\newcommand{\bra}[1]{\ensuremath{\left\langle #1 \right|}}

\begin{document}

\title{\textbf{N\'eel order, ring exchange and
charge fluctuations in the half-filled Hubbard model}}

\author{J.-Y. P. Delannoy}
\affiliation{Laboratoire de Physique, \'Ecole normale sup\'erieure
de Lyon, 46 All\'ee d'Italie, 69364 Lyon cedex 07, France.}

\affiliation{Department of Physics, University of Waterloo,
Ontario, N2L 3G1, Canada}

\author{M. J. P. Gingras}
\affiliation{Department of Physics, University of Waterloo, Ontario, N2L 3G1,
Canada}

\author{P. C. W. Holdsworth}
\affiliation{Laboratoire de Physique, \'Ecole normale
sup\'erieure de Lyon, 46 All\'ee d'Italie, 69364 Lyon cedex 07,
France.}

\author{A.-M. S. Tremblay}

\affiliation{D\'epartement de Physique and RQMP, Universit\'e de
Sherbrooke, Sherbrooke, Qu\'ebec, J1K 2R1, Canada.}

\date{\today}

\begin{abstract}  We  investigate the ground state properties of the two
dimensional half-filled one band Hubbard model in the strong (large-$U$)
to intermediate coupling limit ({\it i.e.} away from the strict Heisenberg limit)
using an effective spin-only low-energy
theory  that includes nearest-neighbor exchange, ring exchange, and all
other  spin  interactions to order $t(t/U)^3$. We show that the operator
for  the  staggered  magnetization, transformed for use in the effective
theory, differs from that for the order parameter of the spin model by a renormalization  factor  accounting  for  the  increased  charge
fluctuations as $t/U$ is increased from the $t/U\rightarrow0$ Heisenberg limit.
These charge fluctuations lead
to an increase of the
quantum fluctuations over and
above   those   for   an   $S=1/2$   antiferromagnet.
The renormalization factor
ensures that the zero temperature staggered moment for the Hubbard model
is  a  monotonously
decreasing function of $t/U$, despite the fact that the moment of the spin Hamiltonien, which depends on transverse spin fluctuations only, in an increasing function of $t/U$. We also comment on quantitative aspects of the $t/U$ and $1/S$ expansions.

\end{abstract}

\maketitle


\section{Introduction}
Effective low-energy theories are constructed and used
in essentially all
fields of Physics. The exponential reduction in the size of the
Hilbert space that occurs in  such theories generally offsets the
disadvantage of working with the non-local operators induced by
elimination of the high-energy states. In the context of
strongly-correlated electrons, spin-only Hamiltonians, such as the
Heisenberg model, are examples of effective low-energy theories
that apply when interactions are very strong. Present research
interests focus on systems in the strong to
intermediate coupling regime,
where one might expect that weaker interactions lead to increased
electron mobility, which in turn should reduce the stability of
magnetic phases. In the effective Hamiltonian, the increased
electron mobility is taken into account perturbatively by
including increasingly higher order corrections to the effective
low-energy theory~\cite{Harris,MacDonald,Chernyshev}. More specifically,
the effective low-energy spin Hamiltonian, $H_{\rm s}$, derived
from the Hubbard model, 
away from the strict Heisenberg limit $[t/U \rightarrow 0]$, contains
conventional Heisenberg pairwise spin exchange as well as
so-called ring (or cyclic) exchange terms that involve $n-$spin
($n>2$) interactions~\cite{MacDonald,taka}. These corrections alter the
low-energy excitations and theoretically they may, if large
enough, produce exotic ground states~\cite{Sandvik}. 
At the present time, there are still many experiments designed  to search for evidence of ring exchange terms in materials such as parent high-temperature superconductors\cite{Toader}.

There exist several methods for deriving effective low-energy
theories such as degenerate perturbation theory, 
canonical transformation, resolvent and projection approaches. 
Their equivalence has been recently demonstrated~\cite{Chernyshev}.
Among the most widely used 
is the so-called canonical transformation
(CT) based on original ideas of van Vleck~\cite{Vleck}.
The main idea behind the CT is the following.
In performing a CT, the ``excursions'' of the
degrees of freedom outside the low energy subspace are taken into account
in the effective theory by non-local effective interactions.
The true ground state eigenvector of the theory is in essence ``rotated''
to lie in the ground state of the subspace of the effective
low energy theory.
The CT method has been extensively used, even outside condensed matter physics.
For example, Foldy and Wouthuysen employed the CT to derive the lowest order
relativistic corrections to the Schr\"odinger equation, starting from the
Dirac equation~\cite{Foldy}.  Two of the best known applications of the CT in
condensed matter physics are the derivation of the Kondo model
from the Anderson impurity model using the Schrieffer-Wolff
CT~\cite{Schrieffer-Wolff} and the derivation, starting from the Hubbard model of an exchange spin Hamiltonian with ring/cyclic exchange terms\cite{Harris,MacDonald,Chernyshev}.  
One important technical aspect 
arising in the construction of effective theories
is that operators defined in the bare high-energy theory must also be canonically transformed
before they can be employed in calculations within the effective low-energy
theory. In the context of condensed matter systems,
the importance of properly transforming operators in high order perturbative
approaches, used to eliminate states from the high energy
sector of the theory, has been emphasized 
in a number of
situations.\cite{Harris,Sawatzky,Eskes:1994,Paramekanti,Hamer,Schmidt}

Consider the Hubbard model with the two energy 
scales defined by $t$ and $U$, where $t$ is the
nearest-neighbor hopping constant and $U$ is the on-site Coulomb energy.
In an effective low-energy theory derived from the Hubbard model and
limited to the spin-only subspace, the
electron hopping processes beyond nearest-neighbor lead to a
4-spin ring exchange term, $J_c$ and to second and third neighbor
exchange interactions, $J_2$ and $J_3$, which are all of order
$(t/U)^2$ smaller than the nearest-neighbor exchange $J_1 \approx
4t^2/U$. Several recent studies have investigated the effect of
$J_c$ on the properties of a $S=1/2$ nearest-neighbor Heisenberg
antiferromagnet~\cite{Lorenzana,Honda,Lauchli,Honda-chain}. In two
dimensions ($2D$) it is found that introducing  a small $J_c$
initially decreases the quantum fluctuations of the N\'eel order
parameter~\cite{Lorenzana,Honda}. Similarly, in a one-dimensional
($1D$) two-leg ladder the spin gap
decreases~\cite{Lauchli,Honda-chain} and the staggered spin$-$spin
correlations increase~\cite{Honda-chain} as $J_c$ is first
increased, again indicating a reduction of quantum fluctuations.
These studies consider $J_c$ as a phenomenological parameter in a
spin model without reference to the microscopic origin of $J_c$
from a Hubbard-like model. However, a tempting interpretation of
the above results is that an increase of $t/U$ away from the
Heisenberg limit increases the N\'eel order parameter,
$M^\dagger$. This picture is re-enforced by a recent
self-consistent Dyson-Maleev spin-wave calculation~\cite{Katanin}
using $H_{\rm s}$ derived from the Hubbard model to order
$t(t/U)^3$ (see Refs.~[\onlinecite{Harris,MacDonald,Chernyshev}]). 
It is found that
$M^\dagger$ for $0< t/U\ll 1$ is increased above the value for the
Heisenberg limit~\cite{Katanin}. This ensemble of results for
effective theories suggests that $M^\dagger$ should pass through a
maximum value at some finite $t/U$ $-$ a conclusion which is
difficult to understand on physical grounds and inconsistent with
Hartree-Fock calculations~\cite{Schrieffer-RPA} and  series expansion~\cite{Shi}, as well as exact diagonalization of the Hubbard model on small clusters~\cite{Fano} where the structure factor measuring staggered magnetic correlations monotonously decreases as $t/U$ increases from the Heisenberg limit.

In this paper, we show that the above paradoxical increase of
$M^\dagger$ as $t/U$ increases in spin-only theories
when compared to calculations
on Hubbard models~\cite{Schrieffer-RPA,Shi,Fano}
is simply resolved if the magnetization
operator is transformed using the unitary transformation,
that eliminates order by order the double
occupancy states in the Hubbard
model~\cite{Harris,MacDonald,Chernyshev,Eskes:1994,Paramekanti}.
As $t/U$ is increased beyond the Heisenberg limit, so that ring-exchange terms 
have to be taken into account, one can identify two quantum corrections to the
antiferromagnetic N\'eel order parameter. The first correction
is an overal amplitude
renormalization factor, coming from short-range charge fluctuations.
The second is the usual transverse spin fluctuations. The
amplitude renormalization factor from charge fluctuations can be
obtained exactly and does not need further calculation, while the
spin fluctuations can be taken into account to a good degree
of approximation using the most naive application of the usual
methods, such as the
$1/S$ Holstein-Primakoff method.
 The $t/U$ dependence of the two effects go in opposite
directions but the amplitude renormalization factor coming from
charge fluctuations dominates, resolving the above paradox and giving
the physically correct trend
of a decreasing N\'eel order parameter as
$t/U$ increases from the Heisenberg limit.
 An analogous result was found in 
Ref.~[\onlinecite{Stephan}] where it was observed that finite $t/U$ 
corrections lead to an increasing Drude weight as $t/U$ increases, 
in contrast to the $t-J$ model, where the weight decreases as $t/U$ increases.

The paper is organized as follows. Starting from the Hubbard Hamiltonian,
we recall in Section IIA the key steps that
lead to an expression of an
effective spin Hamiltonian, based on the CT
method~\cite{Harris,MacDonald}. We then apply this
method in Section IIB
to derive an  expression for the magnetization operator $M^\dagger$
in the low energy, spin only, effective theory. In
 order to expose the quantitative
importance of the charge-fluctuation-induced
renormalization factor acting on $M^\dagger$,
we present in Section III
results from exact diagonalization on small clusters,
and $1/S$ spin wave calculations in the thermodynamic limit.
We end the paper with a brief conclusion in Section IV.
An appendix gives some of the details used in passing from the
fermionic description of the effective theory to the SU(2) spin only
description.

\section{Staggered Magnetization Operator in the Low-Energy Theory}

\subsection{Effective Hamiltonian}

We begin with the one band Hubbard Hamiltonian, $H_{\rm H}$:
\begin{equation}
H_{\rm H}= -t \sum_{i,j;\sigma}
c^\dagger_{i,\sigma}c_{j,\sigma}
+U\sum_i n_{i,\uparrow}n_{i,\downarrow} \;\;\; .
\label{HH}
\end{equation}
The first term is the kinetic energy term that destroys an
electron of spin $\sigma$ at site $i$ and creates it on
nearest-neighbor site $j$. The second term is the on-site Coulomb
interaction : it costs an energy $U$ for two electrons with
opposite spins to remain on the same site $i$; and where
$n_{i,\sigma}=c^\dagger_{i,\sigma}c_{i,\sigma}$ is the number
operator at site $i$. 
We derive the low-energy theory using the canonical transformation method
first used by Harris {\it et al.}~\cite{Harris}
in this context and applied to higher order by MacDonald 
{\it et al.}~\cite{MacDonald}.
The method introduces a
unitary transformation, $e^{i{\cal S}}$, that ``rotates'' $H_{\rm
H}$ into an effective spin-only Hamiltonian, $H_{\rm s}$, and
corresponding state vectors into the restricted spin-only (SO)
subspace.

As introduced in Refs.~[\onlinecite{Harris, MacDonald}], the transformation
relies on the separation of the kinetic part $T$ into three
terms that respectively increase by one ($T_1$), keep ($T_0$)
or decrease by one ($T_{-1}$) the number of doubly occupied sites.
We write:
\eqn{
T\,=\,-t \sum_{i,j;\sigma}
c^\dagger_{i,\sigma}c_{j,\sigma}
 \,=\,T_1\,+\,T_0\,+\,T_{-1}
}
where
\begin{eqnarray}
T_1 &= &-t \sum_{i,j;\sigma}
n_{i,\bar{\sigma}}  c^\dagger_{i,\sigma}c_{j,\sigma} h_{j,\bar{\sigma}} \\
T_{0}&=&-t \sum_{i,j;\sigma}
\!\left ( h_{i,\bar{\sigma}}  c^\dagger_{i,\sigma}c_{j,\sigma} h_{j,\bar{\sigma}}
  + n_{i,\bar{\sigma}}  c^\dagger_{i,\sigma}c_{j,\sigma}
n_{j,\bar{\sigma}}\right) \\
T_{-1}&=&-t \sum_{i,j;\sigma}
 h_{i,\bar{\sigma}}  c^\dagger_{i,\sigma}c_{j,\sigma} n_{j,\bar{\sigma}}
\end{eqnarray}
where $\bar{\sigma}$ stands for up if ${\sigma}$ is down
and for down if $\sigma$ is up.
This decomposition of $T$ comes from right
multiplying the kinetic term in Eq.~(\ref{HH}) by
$n_{i,\bar{\sigma}} + h_{i,\bar{\sigma}} \,=\,1$
and left multiplying by $n_{j,\bar{\sigma}} + h_{j,\bar{\sigma}} \,=\,1$.

Using the Hausdorff formula, the transformation $e^{i{\cal S}}$, applied order by
order in $t/U$ to $H_{\rm H}$, gives:
\begin{equation}
H_{\rm s}\,=\,e^{i {\cal S}}H_{\rm H}e^{-i{\cal S}}
\,=\,H_{\rm H} \,+\, \frac{[i{\cal S},H_{\rm H}]}{1!}\,+\,
\frac{[i{\cal S},[i{\cal S},H_{\rm H}]]}{2!} \,+\, \cdots .
\label{transfo}
\end{equation}
This unitary transformation is chosen so that, to a given order,
the resulting $H_{\rm s}$ does not change the number of doubly-occupied sites. Order by order the weight of the states with double occupancy are reduced. For the complete transformation, $H_{\rm s}$ and the corresponding ground eigenstate vector $|0\rangle_{\rm s}$ are completely confined to the SO subspace.

To third order in $t/U$ and following 
 MacDonald {\it et al.}~\cite{MacDonald} 
we recover the expression of the generator
$i{\cal S}^{(3)}$ of the unitary transformation $e^{i{\cal S}^{(3)}}$ as being:
\begin{eqnarray}
i{\cal S}^{(3)} & = & \f{1}{U}(T_1 - T_{-1}) \+ \f{1}{U^2}([T_1,T_0] - [T_0,T_{-1}])\nonumber\\
& + &  \f{1}{U^3} \left( -\trip{0}{0}{1} - \trip{0}{1}{0} - \trip{1}{1}{0}\right.\nonumber\\
& - & \left. \f{1}{4}\trip{-1}{0}{-1} \+\f{2}{3} \trip{1}{1}{-1}\right.\nonumber\\
& + & \left. \f{2}{3} \trip{-1}{1}{-1} \right)  \;\;\;.
\label{unitary}
\end{eqnarray}
This expression combined with (\ref{transfo}) leads to the expression of the
third order expansion of the Hubbard Hamiltonian in terms of the operators
$T_m$ introduced above. Defining~\cite{MacDonald}
$\displaystyle T^{(k)}(m_1, m_2,\cdots, m_k)
\= T^{k}[m] \= T_{m_1}T_{m_2}\cdots T_{m_k}$,
where $m_k=-1$, 0 or 1, this Hamiltonian reads : 

\begin{eqnarray}
H^{(4)}\,= & - &\f{1}{U} \, T^{(2)}(-1,1)\nonumber \\
& + &\f{1}{U^2} T^{(3)}(-1,0,1) \label{Hs4} \\
& + &\f{1}{U^3} \left(T^{(4)}(-1,1,-1,1) \- T^{(4)}(-1,0,0,1)\right.\nonumber\\
& - &\left.\f{1}{2} T^{(4)}(-1,-1,1,1) \right)\nonumber
\end{eqnarray}

This expression for the effective Hamiltonian needs to be written in a spin only notation. Following Ref.~[\onlinecite{MacDonald}], one can map the singly occupied subspace of states of the Hubbard model onto the states of a Hilbert space of interacting $S=1/2$ spins. The correspondence is:

\eqn{
\begin{array}{ccc}
Hubbard\, space & & Spin\,\f{1}{2} \,space  \\
\\
n_{i,\uparrow} = 1 & \longrightarrow &  \ket{\cdots\underbrace{\uparrow}_{\mbox{site
i}}\cdots}  \\
\\
n_{i,\downarrow} = 1  & \longrightarrow &
\ket{\cdots\underbrace{\downarrow}_{\mbox{site i}}\cdots}
\end{array}
\label{corres}}

The expression of the spin Hamiltonian $H_{\rm s}^{(k)}$, acting on the
spin space, is derived from the Hamiltonian acting on the occupation
number subspace, and is:
\eqn{
H_{\rm s}^{(k)} \=\f{1}{2^N}
\sum_{\alpha_1,\alpha_2,\cdots,\alpha_N = 0}^3 \left(\prod_{l=1}^N
\sigma_{\alpha_l}^{(l)}\right) {\rm Tr}(\sigma_{\alpha_1}^{(1)} \cdots
\sigma_{\alpha_N}^{(N)} H^{(k)}),\label{Tr}
}
where $\sigma_{\alpha_i}^{(p)}$ is the $\alpha_i^{st}$
Pauli matrix associated with site $p$.

A more detailled discussion 
of this mapping is given in Appendix \ref{SH}.
To order $t(t/U)^3$, and dropping constant terms, we recover the results found
using this method in Ref.~[\onlinecite{MacDonald}] and found earlier, via another method~\cite{taka}:
\begin{eqnarray}
H_{\rm s}^{(4)}&=&J_1 \!\! \sum_{<i,j>} {\bf S}_i\cdot{\bf S}_j \,+\, J_2 \!\! \sum_{<i,j_2>}
{\bf S}_i\cdot{\bf S}_{j_2} \,+\, J_3 \!\! \sum_{<i,j_3>}
{\bf S}_i\cdot{\bf S}_{j_3}\nonumber\\
&+&J_c \!\!\!\! \sum_{<i,j,k,l>}
[({\bf S}_i \cdot {\bf S}_j)
({\bf S}_k \cdot {\bf S}_l)+({\bf S}_i \cdot {\bf S}_l)
({\bf S}_j\,\cdot\,{\bf S}_k) \nonumber \\
&& \hspace{43pt}\,-\,  ({\bf S}_i \cdot {\bf S}_k)
({\bf S}_j\,\cdot\,{\bf S}_l)
],
\label{Hso}
\end{eqnarray}
where $j$,  $j_2$ and  $j_3$ are respectively first, second and
third nearest-neighbors of $i$, and $\langle
i,j,k,l\rangle$ denotes the four spins that form an elementary
square plaquette circulating in a clockwise direction. The
coupling constants, homogeneous with an energy $t$, are expanded
to 3$^{\rm rd}$ order polynomials in $t/U$, giving
$J_1=4t^2/U-24t^4/U^3$, $J_2=J_3=4t^4/U^3$, and $J_c=80 t^4/U^3$,
 as in
Ref.~[\onlinecite{MacDonald}].
Since in what follows we only consider $H_s^{(k)}$ to order
$k=4$, we henceforth use $H_s$ as a shorthand notation for
$H_s^{(4)}$.

\subsection{Staggered Magnetization Operator}

The Hubbard ground state wave vector,
$|0\rangle_{\rm H}$, expressed in the effective theory, $e^{i{\cal
S}}|0\rangle_{\rm H}=|0\rangle_{\rm s}$, has a
unique value in the SO subspace.
 However,
it is important to note that $|0\rangle_{\rm s}$ is not simply a
projection of $|0\rangle_{\rm H}$ onto that
space~\cite{Chernyshev}. In performing the transformation the
particle excursions perpendicular to the SO space are taken into account in
the effective theory by the non-local exchange integrals.
The vector $|0\rangle_{\rm H}$ is
therefore rotated by $e^{i{\cal S}}$ to lie entirely in the SO subspace.
Similarly,
physical quantities in the effective theory are not the
expectation values for operators calculated with the projection of
the vectors into the subspace. Since $|0\rangle_{\rm H}=e^{-i{\cal
S}}|0\rangle_{\rm s}$, the expectation value of an operator ${O}_{\rm H}$
in the original Hubbard model can be computed in the
state $|0\rangle_{\rm s}$ as long as the transformed
operator ${O}_{\rm s} = e^{i{\cal S}}  O_{\rm H} e^{-i{\cal S}}$
is used~\cite{Chernyshev,Harris,Sawatzky,Eskes:1994,Paramekanti,Schmidt,Hamer}.
In other words,
\begin{equation}
\left<{O}\right>\,=\,
\frac{_{\rm H}\langle 0|{O}_{\rm H}|0 \rangle_{\rm H}}
{_{\rm H}\langle 0|0 \rangle_{\rm H}}
\,=\,\frac{_{\rm s}\langle 0|{O_{\rm s}}|0\rangle_{\rm s}}
{_{\rm s}\langle 0 |0\rangle_{\rm s}} \;\;\; .
\label{O2}
\end{equation}
These operators ${O}_{\rm s}$ may
differ from the expected form in a phenomenological magnetic model
constructed uniquely in the SO Hilbert space.
We focus here
on the operator for the staggered magnetization (magnetic moment)
 for the Hubbard
model, $M^\dagger_{\rm H}$. We show that, when considered in the
effective theory, the magnetic moment is {\it not the same} as the
Heisenberg magnetic moment operator $\tilde M^\dagger_{\rm s}$. We
henceforth use the tilde symbol  to annotate what an operator,
$\tilde O_{\rm s}$, would be in a SO model with {\it no} relation
to an underlying Hubbard model. We define the conventional
staggered magnetic moment operator, ${M^\dagger _{\rm H}}$, that
lives in the unrestricted Hilbert space of the Hubbard model as
${M^\dagger _{\rm H}} \,=\, (1/N) \sum_i
(n_{i,\uparrow}-n_{i,\downarrow})(-1)^i$. We consider a square
lattice of size $L_x\times L_y=N$, with sites labelled $i \in
[1,N]$.

The unitary  transformation on $M^\dagger_{\rm H}$ ( $M_{\rm s}^{\dag}\,=\,e^{i{\cal S}} M_{\rm H}^{\dag} e^{-i{\cal S}}$ ) is
performed using the Hausdorff formula, as in Eq.~(\ref{transfo}), leading to the expression for a new operator  $M^\dagger_{\rm s}$ in the
SO spin subspace. Such a calculation can  be achieved with the commutation
relations between $M_{\rm H}^{\dag}$ and $T_{-1}$, $T_0$ and $T_1$:

\eqn{[T_1,M_{\rm H}^{\dag}]\,\equiv\,\f{1}{N}\bar{T}_1\,=\,
\f{t}{N} \sum_{i,j,\sigma} (-1)^i \, n_{i,\bar{\sigma}}
c_{i,\sigma}^{\dag} c_{j,\sigma} h_{i,\bar{\sigma}}
\mathbf{\hat{\sigma}}_{\sigma,\sigma}^z,\label{commut1}}

\eqn{[T_{-1},M_{\rm H}^{\dag} ]\,\equiv\,\f{1}{N}\bar{T}_{-1}\,=\,
\f{t}{N} \sum_{i,j,\sigma}(-1)^i\, h_{i,\bar{\sigma}} c_{i,\sigma}^{\dag} c_{j,\sigma}
n_{i,\bar{\sigma}} \mathbf{\hat{\sigma}}_{\sigma,\sigma}^z,\label{commut2}}

\begin{eqnarray}
[T_0,M_{\rm H}^{\dag}] & \equiv &\f{1}{N}\bar{T}_0\,=\,
\f{t}{N} \sum_{i,j,\sigma} (-1)^i\left(n_{i,\bar{\sigma}}
c_{i,\sigma}^{\dag} c_{j,\sigma} n_{i,\bar{\sigma}}
 \mathbf{\hat{\sigma}}_{\sigma,\sigma}^z \right. \nonumber\\
 & +& \left.  h_{i,\bar{\sigma}} c_{i,\sigma}^{\dag} c_{j,\sigma}
h_{i,\bar{\sigma}}  \mathbf{\hat{\sigma}}_{\sigma,\sigma}^z\right).
\label{commut3}
\end{eqnarray}
where we henceforth use the notation $\bar{T}_m$ to emphasize the
similarities between the expression for these new operators and the
original kinetic operators ${T}_m$, and to point out that both
$\bar{T}_m$ and ${T}_m$ increase the number of doubly
occupied sites by $m$.

The spin Hamiltonian in (\ref{Hso}) is the exact expansion of the original microscopic model in (\ref{HH}) obtained by considering terms in ${\cal S}^{(k)}$ ( Eq.(7)~), up to $3^{rd}$ order in $(t/U)$. It is important to realize that an odd power in $t$ cannot appear in the spin expression of $H_s^{(k)}$ ( Eq.(6)~) for any $k$ because we are at half-filling and all states are singly occupied in the low-energy theory. In the case of staggered magnetization, transforming $M_{\rm H}^\dagger$ and retaining terms in ${\cal S}^{(k)}$ up to third order in $(t/U)$ would lead to  a third order power in $t/U$ in the expression of $M_{\rm s}^\dagger$. However, the third order contribution in $t/U$ to $M_{\rm s}^\dagger$ evaluated within the $\vert 0\rangle_{\rm s}$ SO ground state would vanish since we are at half-filling. Hence, we only need to keep terms in ${\cal S}^{(k)}$ up to second order in $(t/U)$ (i.e. the terms proportional to $1/U$ and $1/U^2$). Doing so, we obtain the following expression for the effective staggered magnetization operator in the SO subspace:

\eqn{M_{\rm s}^{\dag}\,=\, M_{\rm H}^\dagger
\+\f{1}{U}\left(\bar{T}_{1} - \bar{T}_{-1}\right)
 \+ \f{1}{2U^2}\left(\bar{T}_{-1}T_1 - T_{-1}\bar{T}_{1}\right)
\label{effmag}}
The linear term only contributes when the expectation of
higher powers of $M_{\rm s}^\dagger$
are computed in the $\vert 0\rangle_{\rm s}$ SO ground state. Then, linear
terms can combine to give an overall $(t/U)^2$ contribution.

Using the same procedure as above in obtaining Eq.~(\ref{Hso}) (see Appendix \ref{SH}) and restricting ourselves to expectation values of the first power of magnetization in the SO subspace, we find the expression for the magnetization operator $M_{\rm s}^\dagger$ in terms of spin $1/2$ operators:

\begin{equation}
{M^\dagger_{\rm s}} = \frac{1}{N}
\left (\sum_i S_i^z(-1)^i
\,-\,2\frac{t^2}{U^2}\sum_{<i,j>} (S_i^z - S_j^z)(-1)^i \right )
\;\;\; .
\label{M3}
\end{equation}
This expression for $M_{\rm s}^\dagger$ contains a correction compared to the standard staggered moment operator $\tilde M_{\rm s}^\dagger$ in a Heisenberg model,
\begin{equation}
\tilde M_{\rm s}^\dagger\,=\, \frac{1}{N}\sum_i S_i^z(-1)^i	\;\;\; .
\end{equation}
This is a consequence of the fact that the original
Hubbard model contains electron
mobility, or charge fluctuations, where particles are allowed to
visit doubly occupied sites. The magnetic moment of the ground
state has therfore non-zero contributions coming from high-energy
configurations with doubly occupied sites. Within the large$-U$
limit, hopping is highly correlated and limited to sequences
taking the system between two configurations in the SO
subspace~\cite{Chernyshev}. When represented in the effective
theory this particle mobility gives rise to additional quantum
fluctuations over and above the quantum spin fluctuations of the
$S=1/2$ spins around a N\'eel ordered state. Hence,  in
calculating the magnetic moment in the effective theory one must
use the operator $M^\dagger_{\rm s}$ and not $\tilde M^\dagger
_{\rm s}$, the latter being used in phenomenological studies
dissociated from a parent high-energy Hubbard-like fermionic
model~\cite{Sandvik,Lorenzana,Honda,Lauchli,Honda-chain,Katanin}.

\subsection{Alternative Derivation}

As an alternative  and possibly more physically transparent 
method to obtain the transformation of the
staggered magnetization $M_{\rm s}^\dagger$ above, and to help
in the physical interpretation of the result,
we add a conjugate field $h_{\rm H}^\dagger $ to the Hubbard staggered
 moment,
\begin{equation}
H_{\rm H}'\,\equiv\, H_{\rm H} \,-\,
h^\dagger_{\rm H}\,\sum_{i} (n_{i,\uparrow}\,-\,n_{i,\downarrow})(-1)^i	\;\;\; ,
\end{equation}
and repeat the unitary transformation calculation starting back at
Eq.~(\ref{transfo}). We find
\begin{eqnarray}
H' _{\rm s} & =&
H_{\rm s}
- h^\dagger_{\rm H} \sum_i \left ( S_i^z(-1)^i -\frac{2t^2}{U^2} \sum_{<j>}
(S_i^z \,-\, S_j^z) (-1)^i
\right )
 \nonumber
\\
&+& 4 (h^\dagger_{\rm H})^2(t^2/U^3) \sum_{<i,j>} {\bf S_i}\cdot {\bf S_j}
\end{eqnarray}
 which satisfies the following relationships
\begin{equation}
\lim_{h^\dagger_{\rm H}\rightarrow 0} - {1\over{N}}\frac{\partial
{H'_{\rm s}}}{\partial h^\dagger_{\rm H}}\,=\, M^\dagger_{\rm s}
\;\;\; \;\;{\rm and}\;\; \lim_{h^\dagger_{\rm H}\rightarrow 0} -
{1\over{N}}\frac{\partial {H'_{\rm s}}}{\partial \tilde
h^\dagger_{\rm s}} \,=\, \tilde M^\dagger_{\rm s} \;\;\; ,
\end{equation}
with $M^\dagger_{\rm s}$ given by Eq.~(\ref{M3})
and where
\begin{equation}\label{heff}
\tilde h^\dagger_{\rm s} \,=\, h^\dagger_{\rm H} \left(1-{2z}\frac{t^2}{U^2}\right), \;
M^\dagger_{\rm s}=\left(1-2z\frac{t^2}{U^2}\right)\tilde M^\dagger_{\rm s},
\end{equation}
with $z$ the nearest-neighbor coordination number. This result gives an alternate point of view (and distinction) of the above relationship between the SO, $\tilde M^\dagger_{\rm s}$,  and Hubbard, $M^\dagger_{\rm H}$, magnetic moments. The moment $\tilde M^\dagger_{\rm s}$ is the response to an effective microscopic conjugate field, $\tilde h^\dagger_{\rm s}$, that is renormalized (reduced) compared with the ``applied'' $h_{\rm H}^\dagger$ staggered  field.
This renormalization of the staggered field
offers another interpretation
for the additional ``amplitude'' fluctuations arising from the
finite electron mobility. Since the weight of doubly
occupied states become more important in the effective theory 
as $t/U$ is increased, 
the local microscopic 
``spin holding'' staggered mean-field $\tilde h_{\rm s}$
as well as $M_s^\dagger(t/U)$ decrease with increasing $t/U$.
As discussed further below, this correction corresponds, to order $(t/U)^2$, to the finite $t/U$ reduction of the spin-density wave amplitude found in the Hartree-Fock solution of the Hubbard model~\cite{Schrieffer-RPA}.

\section{Results and Consequences}
In this section we test the accuacy of the transformation from $M_{\rm H}^\dagger$ to $M_{\rm s}^\dagger$ through exact diagonalisation of small clusters. the behavior of $M_{\rm s}^\dagger$ and $\tilde{M}_{\rm s}^\dagger$ are compared, as a function of $t/U$ in the thermodynamic limit, using spin wave calculations.

\subsection{Behavior of Small Clusters}

As there is no broken symmetry for small systems, we
calculate $M^\dagger_{2,{\rm H}}$ and its SO
counterparts, $M^\dagger_{2,{\rm s}}$ and
$\tilde M^\dagger_{2,{\rm s}}$ defined by

\begin{equation}
M^\dagger_{2,\alpha} =\sqrt{ \langle (M^\dagger _\alpha)^2\rangle} \;\;\;
{\rm and}
\;\;\;
\tilde M^\dagger_{2,{\rm s}} =\sqrt{ \langle (\tilde M^\dagger _{\rm s})^2\rangle}
\;\;\; ,
\end{equation}
where $\alpha \in \{ {\rm H},{\rm s} \}$. (Note that the canonically
transformed $(M^\dagger_{\rm s})^2$ is not the
square of $M^\dagger _{\rm s}$ in Eq.~(\ref{effmag}) \cite{note}).

\begin{figure}[ht]
\begin{center}
\includegraphics[angle=-90,width=0.48\textwidth]{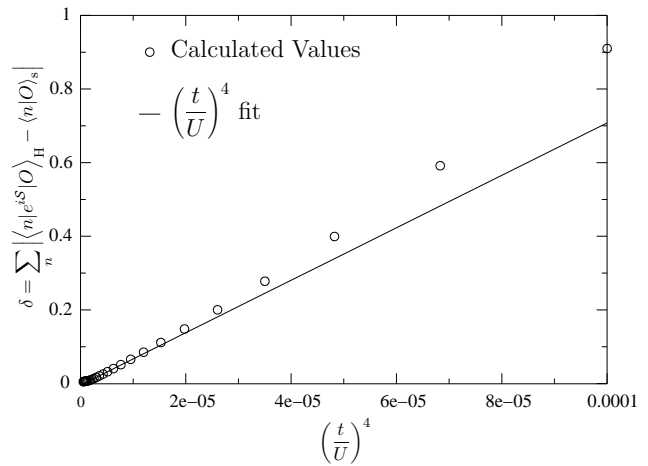}
\caption{
Difference between
$e^{i{\cal S}}\vert 0\rangle_{\rm H}$ and
$\vert 0\rangle_{\rm s}$ restricted to the  singly occupied states.
The result is compared to a $(t/U)^4$ line obtained by fitting $\delta$
in the range $t/U \in [0,0.05]$. The calculation is done on a 
$2\times3$ lattice with open boundary conditions  in both the
$x$ and $y$ directions.
}
\end{center}
\vspace{-7mm}
\end{figure}
For small lattices, of size $L_x \times L_y$, the ground state
$\vert 0\rangle_{\rm H}$ and $\vert 0\rangle_{\rm s}$ of $H_{\rm
H}$ and $H_{\rm s}$ can, respectively, be determined exactly. We
find by direct inspection that the unitary transformation,
$e^{i{\cal S}}$, applied on $\vert 0\rangle_{\rm H}$, indeed
decreases the spectral weight of configurations with doubly
occupied states. As an overall measure of the quantitative
agreement between $e^{i{\cal S}}\vert 0\rangle_{\rm H}$ and $\vert
0\rangle_{\rm s}$ and of the accuracy with which the doubly
occupied states are eliminated from
 $\vert 0\rangle_{\rm H}$,
we plot in Fig. 1
\begin{equation}
\delta\equiv \sum_n \left \vert\, \langle n \vert e^{i{\cal S}} \vert
0\rangle_{\rm H} - \langle n \vert 0\rangle_{\rm s}\, \right \vert
\end{equation}
where the sum is carried over all $2^{(L_xL_y)}$  singly occupied
states. Here a system of size $2\times 3$ with open boundary
conditions
 was considered.
The overlap between the two state vectors diminishes as $t/U$
increases, with a difference that is roughly proportional to
 $(t/U)^4$, the order of the first terms neglected in the calculation.
\begin{figure}[ht]
\begin{center}
\includegraphics[width=0.49\textwidth]{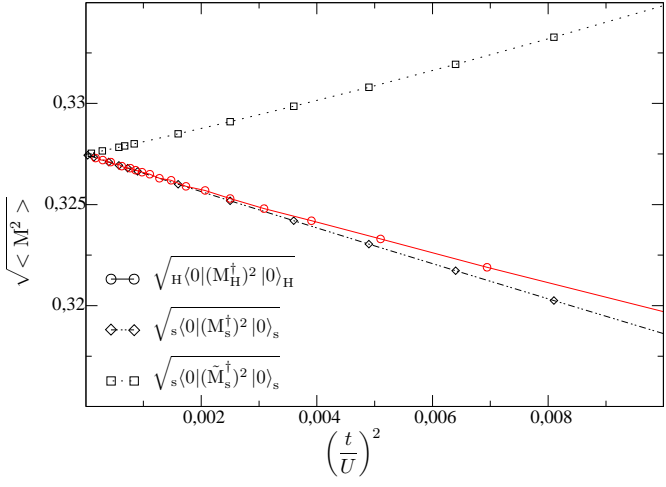}
\caption{
$t/U$ dependence of
the staggered magnetization $M^\dagger_{2,{\rm s}}$,
                           $M^\dagger_{2,{\rm H}}$
 and $\tilde M_{2,{\rm s}}$
for a $2\times 4$ lattice.
}
\end{center}
\vspace{-7mm}
\end{figure}

In Fig. 2 we show results for $M^\dagger_{2,{\rm s}}$,
                  $M^\dagger_{2,{\rm H}}$
and $\tilde M^\dagger _{2,{\rm s}}$ for a $2\times4$ system. The
full curve (circles) shows results for exact diagonalization of
the Hubbard model, $M^\dagger_{2,{\rm H}}$,which should be
considered as the reference data. One can see that $M^\dagger
_{2,{\rm H}}$ is a decreasing function of $t/U$ at small $t/U$, as
expected on physical grounds and as found in previous
exact diagonalizations~\cite{Fano} and
series expansion~\cite{Shi}. The dot-dashed curve (rhombuses) shows
the result for $M^\dagger_{2,\rm s}$. While there is a
quantitative difference between the two results, one finds that
the two sets of data share the same slope, as $(t/U)^2 \rightarrow
0$ and that their difference (not shown) scales as $(t/U)^4$ for
small $t/U$. The dashed curve (squares) shows the $t/U$ dependence
of the magnetic moment calculated from $\tilde M^\dagger _{2,{\rm
s}}$ and $\vert 0\rangle_{\rm s}$. Contrary to the exact result
for $M^\dagger_{2,{\rm H}}$ and the SO result $M^\dagger_{2,{\rm
s}}$, $\tilde M^\dagger _{2,{\rm s}}$ increases with (small)
$t/U$, and never has the correct limiting small $t/U$ behavior.
Simply calculating the staggered magnetic moment, as
defined in a Heisenberg model, is therefore qualitatively
incorrect when the low-energy
Hamiltonian includes higher order corrections in $t/U$. On the
contrary, when the correct SO operator $M^\dagger_{2,{\rm s}}$ is
used, the result is not only qualitatively correct, but the
difference between the exact Hubbard result and the SO result is
less than $1\%$ for $t/U=0.1$.  This suggests that
$(4t/U)^4=.026$, with $4t$ the half-bandwidth, gives an estimate
of the error on the staggered moment in the SO theory. Since the mapping 
between the Hubbard model and the effective theory is not size dependent, 
we expect this accuracy estimate to roughly apply 
in  the thermodynamic limit.

\subsection{Thermodynamic Limit: Spin Wave Calculation}

As, in the absence of boundary effects, $M^\dagger_{\rm s}$ and
$\tilde{M}^\dagger_{\rm s}$ differ only by a multiplicative factor (see
Eq.~(\ref{heff})~), one can estimate the effect of this factor in the
thermodynamic limit within a spin wave approximation. In this case,
the spins operators are written in terms of their bosonic excitations
through a Holstein Primakoff \cite{Holstein,Kittel} $1/S$ expansion
of this bi-partite N\'eel ordered lattice:
\eqn{\begin{tabular}{cc}
Sublattice a  &  Sublattice b \\
$\left\{\begin{tabular}{ccl}
$S_i^z$ & $=$& $S-a_i^{\dagger}a_i$  \\  $S_i^+$&$ =$& $\sqrt{2S}[a_i \! + \!
O(S^{-1})] $ \\ $ S_i^-$ &$ =$& $  \sqrt{2S}[a_i^{\dag}  \!  +  \!  O(S^{-1})]$\end{tabular}\right.$ & $\left\{\begin{tabular}{ccl}
$S_j^z$ & $=$& $-S+b_j^{\dagger}b_j$  \\  $S_j^-$&$ =$& $\sqrt{2S}[b_j \! + \!
O(S^{-1})] $ \\ $ S_j^+$ &$ =$& $  \sqrt{2S}[b_j^{\dag}  \!  +  \!  O(S^{-1})]$\end{tabular}\right.$
\end{tabular}
}

Transforming $H_{\rm s}$ in Eq.~(\ref{Hso}) in reciprocal space, one obtains to
order $S$
\eqn{H\=H_0 \+ \sum_{\vec{k}} A_{\vec{k}}\left(a_{\vec{k}}^{\dag}a_{\vec{k}} \+ b_{\vec{k}}^{\dag}b_{\vec{k}} \right) \+ B_{\vec{k}}\left(a_{\vec{k}}^{\dag}b_{\vec{k}}^{\dag} \+ a_{\vec{k}}b_{\vec{k}} \right),    }
where:

\eqn{\left\{\begin{array}{ccl}
A_{\vec{k}} & = &\displaystyle
4S(J_1-J_2-J_3-J_cS^2 \+
(J_2-J_cS^2) \nu_{\vec{k}} \\& &\displaystyle + J_3\tau_{\vec{k}}) \;\;\; ,\\
B_{\vec{k}} & = &  4S(J_1-J_cS^2) \mu_{\vec{k}} \;\;\; ,\\
H_0 & = & -4NS^2(J_1-J_2-J_3-\f{1}{2}J_cS^2)\end{array}\right. \;\;\;,
}
and
\eqn{\left\{\begin{array}{ccl}
\mu_{\vec{k}} & = & \f{1}{2} (\cos(k_x) \+  \cos(k_y)) \;\;\; , \\
\nu_{\vec{k}} & = & \cos(k_x)\cos(k_y)	\;\;\; , \\
\tau_{\vec{k}} & = & \f{1}{2} (\cos(2k_x) \+  \cos(2k_y)) \;\;\; .
\end{array}\right.
}
Defining
\begin{equation}
\epsilon_{\vec{k}} = \sqrt{A_{\vec{k}}^2 -  B_{\vec{k}}^2}	\;\;\; ,
\end{equation}
we obtain the standard result~\cite{Kittel}
\eqn{
\langle \tilde{M}_{\rm s}^{\dag}\rangle = S - \frac{1}{N} \sum_{\vec{k}}
\left(\frac{A_{\vec{k}}}{\epsilon_{\vec{k}}} - \frac{1}{2}\right)
\;\;\; .
}
We show in Fig. 3
the results for $\langle M^\dagger_{\rm s} \rangle$ and $\langle \tilde
M^\dagger_{\rm s} \rangle$ calculated to order $1/S$ in the
Holstein-Primakoff formulation of the Hamiltonian $H_{\rm s}$ in
Eq.~(\ref{Hso}). The data show qualitatively the same
behaviour as for the exact diagonalization (see Fig. 2): a positive trend
at small $t/U$ for the moment $\tilde M_{\rm s}^\dagger$ of the SO model
and a negative trend for the transformed moment $M_{\rm s}^\dagger$. Even
though the ring exchange term is of order $S^2$ larger than the bilinear
exchange terms, a calculation that would keep boson operators
beyond
quadratic order is apparently not required to get the correct qualitative
trend of $M_{\rm s}^\dagger$ vs $t/U$.
\begin{figure}
\begin{center}
\includegraphics[angle=-90,width=0.45\textwidth]{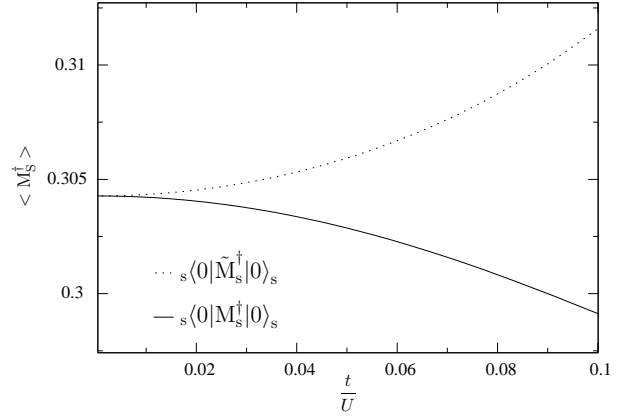}
\caption{
$(t/U)$ dependence of
$\langle M^\dagger_{\rm s} \rangle$ and
 $\langle \tilde M^\dagger_{\rm s} \rangle$
in a Holstein-Primakoff calculation of $H_{\rm S}$ to order $1/S$.
}
\end{center}
\vspace{-7mm}
\end{figure}

These results have several immediate consequences. We conclude
that the increase of the N\'eel order parameter in the presence
of ring-exchange ~\cite{Lorenzana,Honda,Lauchli,Honda-chain,Katanin}
is due to the use of $\tilde
M_{\rm s}^\dagger$, which neglects the renormalization factor
$(1-2zt^2/U^2)$ coming from charge quantum fluctuations. Further, we note that this renormalization factor is, to order $(t/U)^2$,
identical to that
reducing the spin-density wave amplitude in a
Hartree-Fock solution of the Hubbard model~\cite{Schrieffer-RPA}.

We note here that it should not be construed that all quantities
measuring the strength of the magnetic correlations need to be a
monotonously decreasing function of $t/U$. For example, when
considering the three-dimensional Hubbard model, where
the N\'eel temperature $T_{\rm N}$ is nonzero, one finds
that $T_{\rm N}/J =0.9575$ in the Heisenberg limit~\cite{Staudt},
where $J=4t^2/U$ is
the nearest-neighbor Heisenberg exchange. Normalizing
$T_{\rm N}$ by the scale $t$,  we have
$T_{\rm N}/t=3.83 (t/U)$ in the Heisenberg limit~\cite{Staudt,Note_finite_T}.
Hence $T_{\rm N}/t$ is a non-monotonous function of $t/U$, first increasing
as $U$ decreases from the Heisenberg limit, and then decreasing as the weak-coupling limit $t/U \gg 1$ is approached. However, this behavior comes the fact $T_{\rm N}/t$ is controlled by the spin stiffness which scales with $t/U$ in the opposite manner to the zero temperature order parameter, in the strong coupling limit. The spin stiffness is controlled by $J$ but the magnitude of the antiferromagnetic order parameter is not, as can be seen trivially in the Heisenberg limit where it is independent of $J$. More to the point is the observation that
even for a relatively large $U/t=10$,
 $T_{\rm N}/t$
is already 25$\%$ below the N\'eel temperature that would
be predicted by the Heisenberg model~\cite{Staudt}.
Here, the charge fluctuations lower $T_{\rm Neel}$ of the Hubbard model
below that of the corresponding limiting Heisenberg model.
In the context of the work presented here,
it would seem possible
that a numerical calculation on a three-dimensional effective
spin-only Hamiltonian to order $t(t/U)^3$ that neglects charge renormalization
would give a
N\'eel temperature that actually increases even {\it faster}
than the Heisenberg $T_{\rm N}/t=3.83(t/U)$ and definitely faster than $T_{\rm N}$, obtained numerically for the Hubbard model.

\section{Conclusion}

We have shown that transforming the Hubbard model
into an effective spin only theory leads, 
for $t/U$ away from the $t/U=0$ Heisenberg limit,
to a new source of quantum fluctuations that
reduces the staggered magnetization. Indeed, short-range charge
fluctuations renormalize the order parameter by a factor $[1-2z(t/U)^2]$,
depending on $t/U$, which is independent of the spin-only quantum
fluctuations. This factor insures that increasing the charge
mobility reduces the overall stability of the magnetic phase
at $t/U>0$. This is despite the
decrease in long range zero-point spin fluctuations which, when taken alone,
suggests that the antiferromagnetic order parameter should become 
larger as $t/U$ increases from the Heisenberg limit. It would be interesting to 
find out whether this separation of charge and spin
fluctuations is maintained to higher order in the perturbation
scheme.

As a quantitative guide for the validity of the strong-coupling expansion, 
we also checked on small clusters that the difference between 
the result from the Hubbard model and that 
from the spin-only theory is of order $(4t/U)^4$ where the power 
4 is the first power
that is neglected in the derivation of the low-energy theory. 
We also showed that even though the ring exchange term is
of order $S^2$ larger than the bilinear
exchange terms, a calculation that would keep Holstein-Primakoff 
boson operators beyond
quadratic order is apparently not required to get the correct qualitative
trend of $M_{\rm s}^\dagger$ vs $t/U$.

Finally, note that charge fluctuations should also lead to amplitude
renormalization factors for magnetic order at other wave vectors
or for other order parameters such as dimerization.
Renormalization factors for other effective models, such as the
spin model coming from the three band model of the CuO$_2$ plane,
and models that include second, $t'$, and third, $t''$, nearest-neighbor
hopping terms~\cite{DGHT-2},
are also open problems.


\section{Acknowledgements}

We thank
 G. Albinet,
 C. B\'eny,
 E. Dagotto,
 F. Delduc,
 S. Girvin,
 M. Jarrell,
 C. L'Huillier,
 A. L\"auchli,
 J. Lorenzana,
 A. MacDonald,
 A. del Maestro,
 L. Raymond,
 M. Rice,
 R. Scalettar,
 D. S\'en\'echal and
 S. Sorella for useful
discussions. Partial support for this work was provided by NSERC
of Canada and the Canada Research Chair Program (Tier I) (M.G. and
A.T.), Research Corporation and the Province of Ontario (M.G.),
FQRNT Qu\'ebec (A.T.) and a Canada$-$France travel grant from the
French Embassy in Canada (M.G.and P.H.). M.G. and A.T. acknowledge
support from the Canadian Institute for Advanced research.

\appendix
\section{Derivation of the Spin Hamiltonian\label{SH}}
Eq.~(\ref{Hs4}) gives the expression for the
third order expansion of the Hubbard model in terms of $T_m$ operators.
Eqs. (\ref{corres}) and (\ref{Tr}) introduce the mapping between the
spin $S=1/2$  operator Hilbert space and the singly occupied
subspace of the Hubbard model. In this mapping the calculation
of the trace represents a somewhat subtle part because of the
anticommutation relations between the different fermionic operators.
As an example we derive the complete expression of the spin Hamiltonian to the
first non zero order. We start with:
\eqn{H^{(2)} \= -\f{1}{U} \, T^{(2)}(-1,1),}
that is:
\begin{eqnarray}
H^{(2)} & = &\f{t^2}{U} \sum_{i_1,i_2,j_1,j_2}  \left(h_{i_2,\bar{\sigma_2}} c_{i_2,{\sigma_2}}^{\dag} c_{j_2,{\sigma_2}} n_{j_2,\bar{\sigma_2}}\right)\nonumber\\
&& \otimes \left(n_{i_1,\bar{\sigma_1}} c_{i_1,{\sigma_1}}^{\dag} c_{j_1,{\sigma_1}} h_{j_1,\bar{\sigma_1}}\right)\label{h2}.
\end{eqnarray}
Since we work in the singly occupied subspace
($\langle V \rangle = 0$),  the  two electronic
processes that first increase ($T_1$) and
then decrease ($T_{-1}$) the number of
doubly occupied sites have to be performed
between the same 2 sites, which implies for Eq.~(\ref{h2}) that:

\eqn{\left\{\begin{tabular}{ccc}
$j_2$& =& $i_1$\\
$i_2$& =& $j_1$
\end{tabular}\right.}

Defining the fermionic orbitals as:
\eqn{\ket{\underbrace{n_{1\uparrow},n_{1\downarrow}}_{1^{st}\, Site},\,
\underbrace{n_{2\uparrow},n_{2\downarrow}}_{2^{nd}\, Site},\,\cdots
,\,\underbrace{n_{N\uparrow},n_{N\downarrow}}_{N^{th}\, Site}}
\;\;\; .}
For example, we have for two sites:
\eqn{\left\{
\begin{array}{ccccrcr}
c_{1,\uparrow} \ket{\uparrow,\uparrow}& \equiv & c_{1,\uparrow} \ket{1,0,1,0} & = & \ket{0,0,1,0} & \equiv & \ket{0,\uparrow} \\
c_{2,\uparrow} \ket{\uparrow,\uparrow\downarrow}& \equiv & c_{2,\uparrow} \ket{1,0,1,1}  & = & -\ket{1,0,0,1} & \equiv & - \ket{\uparrow,\downarrow}\\
c_{2,\downarrow} \ket{\uparrow,\uparrow\downarrow}& \equiv & c_{2,\downarrow} \ket{1,0,1,1} & = & \ket{1,0,1,0} & \equiv & \ket{\uparrow,\uparrow}
\end{array} \right.,}
the minus sign coming from the odd number of
occupied fermionic orbitals occuring before the one
specified by $c_{i,\sigma}$ or $c_{i,\sigma}^{\dag}$.
We use now the $\uparrow,\downarrow$ notation for simplicity,
but it is important to notice that it represents fermionic
orbital occupancy. The symbol $\ket{0}$ represents a site that is
empty for both its orbitals $\uparrow$ and $\downarrow$.
It follows that:

\eqn{\left\{
\begin{array}{ccc}
T_1 \ket{\downarrow,\uparrow} &=& -t \ket{0,\downarrow\uparrow} \- t\ket{\downarrow\uparrow,0}\\
T_1 \ket{\uparrow,\downarrow} &=& t\ket{0,\downarrow\uparrow} \+ t\ket{\downarrow\uparrow,0}
\end{array} \right.,}

\eqn{\left\{
\begin{array}{ccc}
T_{-1} \ket{0,\downarrow\uparrow} &=& -t\ket{\downarrow, \uparrow} \+ t\ket{\uparrow,\downarrow}\\
T_{-1} \ket{\uparrow\downarrow,0} &=& -t\ket{\downarrow, \uparrow} \+ t\ket{\uparrow,\downarrow}
\end{array} \right.,}

and finally:

\eqn{\left\{
\begin{array}{ccc}
T_{-1}T_1 \ket{\uparrow,\downarrow} & = & 2t^2\left\{  \ket{\uparrow,\downarrow} \- \ket{\downarrow, \uparrow} \right\} \\
T_{-1}T_1 \ket{\downarrow,\uparrow} & = & -2t^2\left\{  \ket{\uparrow,\downarrow}\ - \ket{\downarrow, \uparrow} \right\}
\end{array} \right.	\;\;\; .}

We can then rewrite (\ref{h2}) as:
\eqn{H^{(2)} \= -\f{t^2}{U} \left( \ket{\uparrow,\downarrow}\ - \ket{\downarrow, \uparrow}\right) \left( \bra{\uparrow,\downarrow}\ - \bra{\downarrow, \uparrow}\right)}
This form makes it easier to calculate the trace in (\ref{Tr}) and gives:

\eqn{{\rm Tr}(\sigma_1^x\sigma_2^x H^{(2)})\=  -2 \f{t^2}{U}}
\eqn{{\rm Tr}(\sigma_1^y\sigma_2^y H^{(2)})\=  -2 \f{t^2}{U}}
\eqn{{\rm Tr}(\sigma_1^z\sigma_2^z H^{(2)})\=  -2 \f{t^2}{U}}
\eqn{{\rm Tr}(\sigma_1^0\sigma_2^0 H^{(2)})\=  2 \f{t^2}{U}}

so that:
\eqn{H_{\rm s}^{(2)} \= -\f{t^2}{U} \sum_{<i,j>}
\left( 1 - \bf{\sigma_i}  \cdot \bf{\sigma_j} \right)	\;\;\;	, }
or
\eqn{H_{\rm s}^{(2)} \= -\f{4t^2}{U} \sum_{<i,j>}\left( \f{1}{4} - \bf{S_i}  \cdot \bf{S_j} \right)	\;\;\;	, }
recovering the well known result of the nearest neighbor
interaction coupling constant $J_1$:
\eqn{J_1\=\f{4t^2}{U}}
The same method can be applied to calculate the expression of the
spin Hamiltonian up to order $t(t/U)^3$. 
As in Ref.~\onlinecite{MacDonald}, we used a program
for the general construction of $H_s^{(k)}$ for $k\ge 4$.
To order $k=4$, this leads to Eq.~(\ref{Hso}).
\begin{eqnarray}
H_{\rm s}^{(4)}
&=&J_1 \!\! \sum_{<i,j>} {\bf S}_i\cdot{\bf S}_j \,+\, J_2 \!\! \sum_{<i,j_2>}
{\bf S}_i\cdot{\bf S}_{j_2} \,+\, J_3 \!\! \sum_{<i,j_3>}
{\bf S}_i\cdot{\bf S}_{j_3}\nonumber\\
&+&J_c \!\!\!\! \sum_{<i,j,k,l>}
[({\bf S}_i \cdot {\bf S}_j)
({\bf S}_k \cdot {\bf S}_l)+({\bf S}_i \cdot {\bf S}_l)
({\bf S}_j\,\cdot\,{\bf S}_k) \nonumber \\
&& \hspace{43pt}\,-\,  ({\bf S}_i \cdot {\bf S}_k)
({\bf S}_j\,\cdot\,{\bf S}_l)
],
\end{eqnarray}

\vspace{-6mm}



\begin{thebibliography}{99}
\bibitem{Note_finite_T} This means that in the large
$U/t$ limit, at some finite temperature $T$,
antiferromagnetic correlations must increase with increasing $t/U$, going from
zero staggered magnetization to a finite value as one crosses the phase
boundary.
(See Ref.~[\onlinecite{Staudt}]).
\end{thebibliography}
\end{document}